\documentclass[11pt,twoside,onecolumn]{article}
\usepackage[]{amsmath,amssymb}
\usepackage[]{epsfig}
\newcommand{\be}{\begin{eqnarray}}
\newcommand{\ee}{\end{eqnarray}}

\newcommand{\expec}[1]{\mbox{$\langle\, #1\,\rangle$}}

\renewcommand{\d}{\mbox{{\rm d}}}
\title{\bf Radion Induced Spontaneous Baryogenesis}
\author{G.L.~Alberghi\thanks{e-mail: alberghi@bo.infn.it},$\ $
R.~Casadio\thanks{e-mail: casadio@bo.infn.it}
$\ $ and A.~Tronconi\thanks{e-mail: tronconi@bo.infn.it}\\
 \\
{\em Dipartimento di Fisica, Universit\`a di Bologna, and}
\\
{\em Istituto Nazionale di Fisica Nucleare,
Sezione di Bologna, Italy}}
\begin{document}
%
%
\maketitle
\begin{abstract}
\noindent
We describe a possible scenario for spontaneous baryogenesis given by
the Randall-Sundrum brane-world model with two branes and a radion stabilizing potential.
It is known that the cosmological expansion of the Universe can be recovered
in this brane-world model by adding matter on the branes, and time-reversal is then broken.
We show here that, as a consequence of such a time dependence of the system,
the effective coupling of the radion with matter fields generates a chemical potential
with opposite signs for baryons and anti-baryons and spontaneous baryogenesis is therefore
induced naturally and self-consistently.
\end{abstract}
%
\pagestyle{plain}
\raggedbottom
\setcounter{page}{1}
\section{Introduction}
\setcounter{equation}{0}
\label{intro}
One of the most peculiar features of our Universe is the observed baryonic
asymmetry.
This can be conveniently characterized by the dimensionless number
\be
{n_B\over s} \equiv \eta
\simeq 10^{-10}
\ ,
\label{eta}
\ee
where $n_B \equiv n_b - n_{\bar b}$ is the difference between the
baryon and anti-baryon densities and $s$ is the density of entropy.
The consistency of primordial nucleosynthesis, which yields some of the most
precise results in the standard model of cosmology, requires that $ \eta $
took the above value at the time when the light elements
({\em i.e.},~$^3$He, $^4$He, and $^7$Li) were produced,
and it is believed to have then remained the same up to the present epoch.
\par
The necessary conditions for generating the baryonic asymmetry in quantum
field theory were formulated by Sacharov in 1967~\cite{Sakharov} (see also
Ref.~\cite{DimopoulosSusskind}) and can be summarized as follows:
\begin{enumerate}
\item
Different interactions for particles and antiparticles, or, in other words,
a violation of the C and CP symmetries;
\item
Non-conservation of the baryonic charge;
\item
Departure from thermal equilibrium.
\end{enumerate}
The last condition results from an application of the CPT theorem~\cite{cpt1,cpt2}.
In fact, CPT~invariance of quantum field theory in a static Minkowski space-time
ensures that the energy spectra for baryons and anti-baryons are identical,
leading consequently to identical distributions at thermal equilibrium.
This explains why the baryon number asymmetry was required to be generated
out of thermal equilibrium.
\par
The so called mechanism of spontaneous baryogenesis~\cite{CohenKaplan1,CohenKaplan2}
uses the natural (strong) CPT non-invariance of the Universe during its early history
to bypass this third condition.
We know that an expanding Universe at finite temperature violates both
Lorentz invariance and time reversal, and this can lead to effective CPT
violating interactions~\cite{cpt1,cpt2}.
Thus the cosmological expansion of the early Universe leads us naturally
to examine the possibility of generating the baryon asymmetry in thermal
equilibrium.
The main ingredient for implementing this mechanism is a scalar field $\phi$
with a derivative coupling to the baryonic current.
If the current is not conserved and the time derivative of the scalar
field  has a non-vanishing expectation value, an effective chemical
potential with opposite signs for baryons and anti-baryons is generated
leading to an asymmetry even in thermal equilibrium.
\par
The brane-world model with two branes proposed by Randall
and Sundrum (RS) in Ref.~\cite{Randall} contains a metric degree of freedom
called the {\em radion\/} which determines the distance between the two branes
and appears as a scalar field $\phi$ on the branes.
Cosmological solutions have also been examined rather extensively in this
context.
In particular, it has been shown that, when matter is added on one (or
both) of the two branes, the standard Friedmann equation for the scale
factor of the Universe is recovered (with possible corrections) provided
the radion is suitably stabilized (see for example
Refs.~\cite{RandallCsaki}-\cite{radyon3} and References therein).
In this brane-world model, we therefore have both a scalar field (the radion) and
the cosmological evolution as required by spontaneous baryogenesis,
and we shall show that the radion field does in fact couple differently
with baryons and anti-baryons.
This scenario might therefore naturally reproduce the observed baryonic
asymmetry~\footnote{Other mechanisms for baryogenesis
in the context of brane-world models have been recently analyzed,
e.g.~in Refs.~\cite{martin1}-\cite{martin7}.}.
\par
In Section~\ref{SPsection}, we review in some details the mechanism
of spontaneous baryogenesis.
The cosmological solutions in the RS framework are discussed
in Section~\ref{RSsection} where the process of spontaneous baryogenesis
driven by the radion is presented in general.
Some more specific examples are also reported in Section~\ref{ex}.
We then conclude and comment on our results.
\par
We shall use units with $c=\hbar = k_{\rm B} =1$, where $k_{\rm B}$ is the Boltzmann
constant.
\section{Spontaneous Baryogenesis}
\setcounter{equation}{0}
\label{SPsection}
To illustrate the mechanism of spontaneous baryogenesis
(see, {\em e.g.},~Refs.~\cite{CohenKaplan1,CohenKaplan2}
and \cite{Trodden}-\cite{china2}) let us consider
a theory in which a neutral scalar field $\phi$ is coupled to the baryonic
current $ J^{\mu}_B $ according to the Lagrangian density
\be
L_{\rm int} = \frac{\lambda'}{M_{\rm c} }\, J^{\mu}_B\,\partial _{\mu} \phi
\ ,
\ee
where $\lambda '$ is a coupling constant and $M_c$ is a cut-off mass scale
in the theory (presumably smaller than the Planck mass $M_{\rm Pl} $).
Let us assume that $ \phi $ is homogeneous, so that only the time
derivative term contributes,
\be
L_{\rm int} = \frac{ \lambda '}{M_{\rm c} }\,\dot \phi\, n_B
\equiv
\mu(t)\,  n_B
\ ,
\label{Lint}
\ee
where $ n_B = J^0_B $ is the baryon number density and $ \mu(t) $ is to
be regarded as an effective time-dependent chemical potential.
This interpretation (see Ref.~\cite{Dolgov}) is valid if the current
$ J^{\mu}_B $ is not conserved (otherwise one could integrate the
interaction term away) and if $ \phi $ behaves as an external field which
develops a slowly varying time derivative $ \expec{\dot \phi} \neq 0 $
as the Universe expands.
Since the chemical potential $\mu$ enters with opposite signs for baryons
and anti-baryons, we have a net baryonic charge density in thermal equilibrium
at the temperature $T$,
\be
n_B (T; \xi )
=
\int \frac{\d^3 k}{(2 \pi)^3 }\,\left[ f(k,\mu) - f(k,-\mu) \right]
\ ,
\ee
where $ \xi \equiv \mu / T $ is regarded as a parameter, and
\be
f(k, \mu) =
\frac{1}{\exp \left[ \left( \sqrt{k^2 + m^2} - \mu \right) /T \right]
\pm 1}
\label{distrib}
\ee
is the phase-space thermal distribution~\footnote{Of course, the plus sign is for
fermions and the minus sign for bosons.}
for particles with rest mass $m$ and momentum $ k $.
For $|\xi|\ll 1 $ we may expand Eq.~(\ref{distrib}) in powers of
$ \xi $  to obtain
\be
n_B (T; \mu)=\frac{g\,T^3}{6}\,\xi + O\left( \xi ^2\right)
\ ,
\ee
where $g$ is the number of degrees of freedom of the field corresponding
to $n_B$.
Upon substituting in for the expression of $\mu$, one therefore finds
\be
n_B (T;\mu)
\simeq
\frac{ \lambda'\, g}{6\,M_{\rm c} }\, T^2\,\expec{\dot\phi}
\ .
\label{asymmetry}
\ee
\par
Regardless of the specific mechanisms which break baryon number conservation,
we assume that there is a
temperature $T_{\rm F}$ at which the baryon number violating processes become
sufficiently rare so that $n_B$ freezes out ($T_{\rm F}$ will in fact be called the
{\em freezing temperature\/}).
Once this temperature is reached as the universe cools down, one is left
with a baryonic asymmetry whose value is given by Eq.~(\ref{asymmetry})
evaluated at $T=T_{\rm F}$.
The value of the parameter $ \eta $ remains unchanged in the subsequent
evolution.
\section{Radion Induced Spontaneous Baryogenesis}
\setcounter{equation}{0}
\label{RSsection}
We have discussed how the mechanism of spontaneous baryogenesis may explain
the observed baryonic asymmetry.
We shall now argue that it might occur naturally in brane-world models.
In particular, we shall consider the five-dimensional RS model of Ref.~\cite{Randall}
perturbed by matter on one or both branes~\cite{radyon1}-\cite{radyon3}.
The reader is referred to Ref.~\cite{RandallCsaki} for more details on the
framework and notation used hereafter.
\par
In this model the metric can be written in the form
\be
\d s^2
&=&
n^2(y,t)\,\d t^2
- a^2(y,t)\,\left[\left(\d x^1\right)^2 + \left(\d x^2\right)^2
+ \left(\d x^3\right)^2\right]
- b^2(y,t)\,\d y^2
\nonumber
\\
&\equiv&
\tilde g_{AB}(x,y)\,\d x^A\,\d x^B
\ ,
\ee
where $t$ is the time, $x^i$ are the spatial coordinates along the branes
and $y$ is the extra-dimensional coordinate.
In this formalism, the {\em Planck\/} brane is conventionally located at $y=0$
and the {\em TeV\/} brane at $y=1/2$.
The Einstein equations are given by $G_{AB} = \kappa^2\,T_{AB}$,
where $ \kappa ^2 = 1/(2\,M^3)$ and  $M$  is the five-dimensional Planck
mass.
The energy-momentum tensor $T_{AB}$ contains a contribution
from the bulk cosmological constant $\Lambda$ of the form
$T_{AB}^{\rm bulk} = \Lambda\,\tilde g_{AB}$
and a contribution from the matter on the two branes,
\be
T_A^{\ B\ {\rm branes}}
&=&
\frac{1}{b }\,\delta(y)\,
{\rm diag}\left[V_{*}+\rho_{*}, V_{*}-p_{*}, V_{*}-p_{*}, V_{*}-p_{*}, 0
\right]
\nonumber
\\
&&
+\frac{1}{b}\,
\delta (y-1/2)\,{\rm diag}
\left[V+\rho, V-p, V-p, V-p, 0\right]
\ ,
\ee
where $V_{*}$ is the (positive) tension of the Planck brane and
$V$ the (negative) tension on the TeV brane.
We have correspondingly denoted by $ \rho_{*}$ and $p_{*}$ the density
and pressure of the matter localized on the positive tension (Planck)
brane (assuming an equation of state of the form
$p_{*} = w_{*}\,\rho_{*}$)
and by $\rho$ and $p$ the density and pressure of the matter on the
negative tension (TeV) brane.
Once a stabilizing potential for the radion is included, the stress-energy
tensor picks up an additional term and the solution of the
Einstein equations may be written as a perturbation of the usual
RS solution,
\be
n(y) = a(y) = e^{- m_0\,b_0\,|y|}
\ ,
\ee
with $V_{*}={6\,m_0 / \kappa^2}=-V$ and
$\Lambda=-{6\,m_0^2 / \kappa }$.
We also recall that the constants $b_0$ and $m_0$ determine the effective
four-dimensional Planck mass as
$(8\,\pi\,G_N )^{-1}= M_{\rm Pl}^2=(1 - \Omega_0^2)/ \kappa^2\,m_0$,
where $ \Omega_0 \equiv  e^{- m_0\,b_0/2 } $.
\par
In order to obtain an effective action for the four-dimensional theory,
one perturbs the metric about the RS solution in the form
\be
a(t,y)
&=&
a(t)\, e^{- m_0\, b(t)\, |y|} [ 1 + \delta a(y,t) ]
\nonumber
\\
\nonumber
\\
n(t,y)
&=&
e^{- m_0\, b(t)\, |y|}  [ 1 + \delta n(y,t) ]
\\
\nonumber
\\
b(t,y)
&=&
b(t)  [ 1 + \delta b(y,t) ]
\ ,
\nonumber
\ee
and drops the metric perturbations which contribute only to second order
in $ \delta a$, $ \delta n $ and $ \delta b $ (see Refs.~\cite{radyon1}-\cite{radyon3}
for the effects
of the latter).
It is then useful to introduce the notation
$ \Omega(y,b(t)) = e^{- m_0\,b(t)\,|y|} $ and
$\Omega_b \equiv \Omega(1/2, b(t))$
($\Omega _b$ evaluated at $b = b_0$ is then given by $\Omega_0$).
By integrating over the fifth dimension, one obtains an effective
action for the radion field.
Further, upon examining the equations of motion for $b(t)$, one notes that,
since $\Omega_b$ depends on $b$, the presence of matter on
the two branes generates an effective potential for $b(t)$ given by
\be
V _{\rm eff}( b )
=
V_r(b)
+\frac{ f^4(b)}{4 }\,
\left[ \rho_{*} -3\,p_{*}+ \left(\rho-3\,p\right) \Omega_b^4 \right]
\ ,
\ee
with
\be
f(b) =  \left(\frac{1-\Omega_0^2}{1-\Omega_b^2} \right) ^{1/2}
\ .
\ee
The function $V_r=V_r(b(t))$ is the potential which would stabilize the
radion at the value $b=b_0$ in the absence of matter.
It can therefore be expanded near its minimum
as~\footnote{This expression follows
from Eq.~(4.12) of Ref.~\cite{RandallCsaki} by defining
$\sqrt{3/2}\, \phi / \Lambda_W = m_0\,b $
(where $\Lambda_W=\Omega_0\,M_{\rm Pl}\simeq 1\,$TeV).}
\be
V_r (b) = \frac{1}{4}\, m_r^2\,
\left(\frac{ m_0\, b_0}{1- \Omega_0^2} \right)^2\,
 \Omega_0^2\, M_{\rm Pl}^2\, \left(\frac{b-b_0}{b_0} \right) ^2
\ ,
\ee
where
\be
m_r^2 
=
\frac{ 2 \left(f^4\,V_r\right)'' (b_0)}{3\, m_0^2\,M_{\rm Pl}^2\,\Omega_0 ^2 }\,
\left( 1- \Omega_0^2\right)^2
\ ,
\ee
and $m_r$ is the effective radion mass.
Thus the radion, in the presence of matter on the two branes,
is stabilized to a shifted value $ b_0 + \Delta b$ determined by
\be
{ \Delta b \over b_0} =
\frac{1}{3 }\left(\frac{ 1- \Omega_0^2}{m_0\, b_0}\right)
\frac{ \rho - 3 p  + \Omega_0^2 \left(\rho _{*}  -3 p_{*}\right)}
{m_r^2\,\Omega_0^2\,M_{\rm Pl}^2}
\ ,
\label{deltab}
\ee
where $\Delta b$ is the distance between the minima of
$V_{\rm eff}$ with and without matter.
\par
From the above expression, we see that, since the trace of the stress-energy
tensor vanishes for radiation, even if the Universe on the TeV brane is in
a radiation dominated era ($\rho\simeq 3\,p$), the radion evolution is determined
by the behavior of massive matter on the TeV and Planck branes.
This is the essential ingredient which allows for the possibility of
inducing spontaneous baryogenesis.
We note that the factor $\Omega_0^2 $ in front of $ \rho_{*} $
would make this term negligible for comparable energy densities on
the TeV and Planck branes, but
the fact that the natural energy scale on the Planck brane is of
the order of $M_{\rm Pl} $ may nonetheless allow for a relevant
contribution to baryogenesis from the Planck brane.
\par
Let us now assume that the high energy Lagrangian for matter
on the TeV brane contains an interaction term of the form given in
Eq.~(\ref{Lint}),
\be
L_{\rm int} =  \lambda'\, m_0\, \dot b\, n_B
\ ,
\ee
where $\dot b$ now plays the role of $\expec{\dot\phi}$.
Such a term is the same as that in Eq.~(4.30) of Ref.~\cite{RandallCsaki}.
On using Eq.~(\ref{deltab}) to estimate $ \dot b\simeq \dot{\Delta b} $,
we finally obtain
\be
\eta =\frac{m_0\,\dot b}{T }
=  
\frac{1}{m_r^2\,\Omega_0^2\,M_{\rm Pl}^2 }
\left(\frac{1}{T}\right)\frac{\d}{\d t} \left[\left( \rho - 3\,p\right)
+ \Omega_0^2\,\left( \rho_{*} - 3\,p_{*} \right) \right]
\ . 
\label{bdot}   
\ee
Due to the expansion of the Universe, the time derivative on the right hand side
of this equation will in general acquire a non vanishing (expectation) value
and the baryonic symmetry is therefore dynamically broken in the model.
We also note that the radion field is likely very massive~\footnote{For
example, if the radion is stabilized by the Goldberger-Wise
mechanism~\cite{GoldWise}, one has $ m_r \simeq 1\,$TeV.}, and one can
then assume that the radion follows instantaneously any changes
of the matter density.
\section{Examples}
\setcounter{equation}{0}
\label{ex}
In order to complete our analysis, we shall now estimate the baryonic
asymmetry~(\ref{bdot})
at the freezing temperature $T_{\rm F}$ in three specific scenarios:
\begin{enumerate}
\item
\label{I}
If the effect of matter on the Planck brane is negligible in Eq.~(\ref{bdot}),
the condition to generate the observed baryonic asymmetry~(\ref{eta}) can be
estimated as 
\be
\left. \frac{1}{T }\,\frac{\d}{\d t} \left( \rho - 3\,p\right) \right| _{T_{\rm F}}
\gtrsim 10 ^{-10}\,{\rm TeV}^4
\ .  
\ee
Let $ \rho_m $ be the energy density of any non-traceless component of the
energy momentum tensor.
By using the continuity equation and the Friedmann equation for a radiation
dominated Universe up to an energy density of the order of  $1\,$TeV$^4$,
 which is roughly the limit  of validity for the RS model,
one obtains the requirement
\be
\rho_{m} \gtrsim 10^{-6}\,{\rm TeV}^4
\ee
for this mechanism to produce sufficient baryonic asymmetry.
\item
\label{II}
Since the Planck brane remains hidden, one can allow
for a very large term proportional to $\dot\rho_*$ in Eq.~(\ref{bdot})
(we recall that matter energy on the Planck brane is allowed
up to the Planck scale).
The radion velocity would hence be larger than in the previous case,
and the freezing temperature correspondingly lower.
The bound in this case is
\be
\left.\frac{1}{T }\,\frac{\d}{\d t}\left( \rho_{*} - 3\,p_{*}\right) \right| _{T_{\rm F}}
\gtrsim 10 ^{-42}\,M_{ \rm Pl}^4
\ .
\ee
\item
\label{III}
Another possibility is to consider the stage when the radion is still stabilizing
towards the equilibrium value $ b_0 $ and the effect of matter on the branes
is negligible.
A typical radion velocity would be larger than in both previous cases,
$  m_0\,\dot b \simeq H(T)\,M_{\rm Pl} / \Lambda_W $, and 
\be
T\gtrsim 10^{2}\, {\rm eV}
\ , 
\ee
which allows the widest range of temperature among the three possibilities outlined.
\end{enumerate}
Of course, the above list is not exhaustive and one could consider many
other situations.
For example, one could include more bulk fields or different
couplings between the radion and brane fields.
A complete analysis of all possible cases however goes beyond the scope of the
present work and will not be given here.  
\section{Conclusions}
\setcounter{equation}{0}
\label{conc}
We have shown that the perturbations induced by the addition of matter
on one (or both) of the two branes of a cosmological RS model with
a stabilizing potential for the radion naturally lead to a non-vanishing
expectation value for the velocity of the radion field.
Since the latter couples with the baryonic current on the branes,
this naturally induces the onset of spontaneous baryongenesis, as described
by the general formula~(\ref{bdot}).
\par
Having outlined the main ideas in the present paper, the next step would be to
analyze all possible scenarios.
For specific cases, it may in fact be possible to reproduce the observed baryonic
asymmetry $\eta$ in Eq.~(\ref{eta}).
Conversely, the required value of $\eta$ can be viewed as a constraint that
brane-world models must satisfy.
\section*{Acknowledgments}
R.C.~would like to thank G.~Lambiase for comments.
\end{document}